\documentclass[numerical,twocolumn,superscriptaddress,showpacs,floatfix,reprint]{revtex4}

\usepackage{graphicx,color}
\usepackage{amsmath,amssymb,amsfonts}
\usepackage[colorlinks=true,plainpages=true]{hyperref}

\def \be {\begin{equation}}
\def \ee {\end{equation}}
\def \ee  {\end{equation}}
\def \bea {\begin{eqnarray}}
\def \eea {\end{eqnarray}}

\begin{document}
\title{
A study of nuclear structure of light nuclei at the Electron-Ion Collider
}

\medskip

\author{Niseem~Magdy} 
\email{niseemm@gmail.com}
\affiliation{Department of Chemistry,  Stony Brook University, Stony Brook, NY 11794, USA}
\affiliation{Center for Frontiers in Nuclear Science at SBU, Stony Brook, NY 11794, USA}
\affiliation{Department of Physics, University of Tennessee, Knoxville, TN, 37996, USA}

\author{Mariam Hegazy}
\affiliation{Department of Physics, Faculty of Science, Cairo University, Giza 12613, Egypt}

\author{Aliaa Rafaat}
\affiliation{Department of Physics, The American University in Cairo, New Cairo 11835, Egypt}

\author{Wenliang~Li}
\email{billlee@jlab.org}
\affiliation{Department of Physics, Stony Brook University, Stony Brook, NY 11794, USA}
\affiliation{Center for Frontiers in Nuclear Science at SBU, Stony Brook, NY 11794, USA}

\author{Abhay~Deshpande} 
\affiliation{Department of Physics, Stony Brook University, Stony Brook, NY 11794, USA}
\affiliation{Department of Physics, Brookhaven National Laboratory, Upton, New York 11973, USA}
\affiliation{Center for Frontiers in Nuclear Science at SBU, Stony Brook, NY 11794, USA}

\author{A. M. H. Abdelhady}
\affiliation{Department of Physics, Faculty of Science, Cairo University, Giza 12613, Egypt}

\author{A.Y.Ellithi}
\affiliation{Department of Physics, Faculty of Science, Cairo University, Giza 12613, Egypt}

\author{Roy~A.~Lacey}
\affiliation{Department of Chemistry,  Stony Brook University, Stony Brook, NY 11794, USA}
\affiliation{Department of Physics,    Stony Brook University, Stony Brook, NY 11794, USA}

\author{Zhoudunming~Tu} 
\email{zhoudunming@bnl.gov}
\affiliation{Department of Physics, Brookhaven National Laboratory, Upton, New York 11973, USA}


\begin{abstract}
{\color{black} Understanding the substructure of atomic nuclei, particularly the clustering of nucleons inside them, is essential for comprehending nuclear dynamics.} Various cluster configurations can emerge depending on excitation energy, the number and types of core clusters, and the presence of excess neutrons. Despite the prevalence of tightly bound cluster formations in low-lying states, understanding the correlation between clusters and their formation mechanisms remains incomplete. This exploring study investigates nuclear clustering at the Electron-Ion Collider (EIC) using simulations based on the modified BeAGLE model. By simulating collisions involving $e$+$^{9}$Be, $e$+$^{12}$C, and $e$+$^{16}$O nuclei, we find that the average energy of particles $\langle E \rangle$ and the system size ratios of particles at forward rapidity exhibit sensitivity to alpha clustering and its various configurations. These findings offer valuable insights into the dynamics of nuclear clustering and its implications for future studies at the EIC.
\end{abstract}
\maketitle
\section{Introduction}
Exploring the complex interactions among nucleons within atomic nuclei remains a central challenge in nuclear physics, driving research into diverse theoretical frameworks and experimental methodologies. Of particular interest is the inquiry into cluster structures within nuclear matter, which sheds light on the intricate interplay between nucleons and the emergence of distinctive structural motifs~\cite{Freer:2017gip, AFZAL:1969iik, Feldmeier:2000cn, Lee:2008fa, Roth:2010bm, Schuck:2016fex, Tohsaki:2017hen}.

{\color{black}
Cluster structures arise when nucleons aggregate into discernible groups, presenting diverse shapes and sizes. 
In principle, the study of nuclear clustering began with Rutherford's discovery of alpha radiation~\cite{Rutherford:1899} and the advent of quantum mechanics. In 1928, Gamow \cite{Gamow:1928zz}, along with Gurney and Condon~\cite{Gurney:1929zz}, described the $\alpha$-particle as undergoing quantum-mechanical tunneling from within the decaying nucleus. About a decade later, in 1937, Wheeler \cite{Wheeler:1937zz} developed the resonating group method to describe $\alpha$-clusters and other cluster formations within nuclei while preserving the fermionic quantum statistics of protons and neutrons.}
Later, Ikeda's seminal work in 1968 \cite{Ikeda:1968} introduced the Ikeda diagram, enriching our comprehension by predicting various cluster structures within light nuclei. Notably, mean-field effects have been found insufficient to disrupt cluster formation \cite{He:2016cwt}.
 \begin{figure}[!h]
 \centering{
 \includegraphics[width=0.9\linewidth,angle=0]{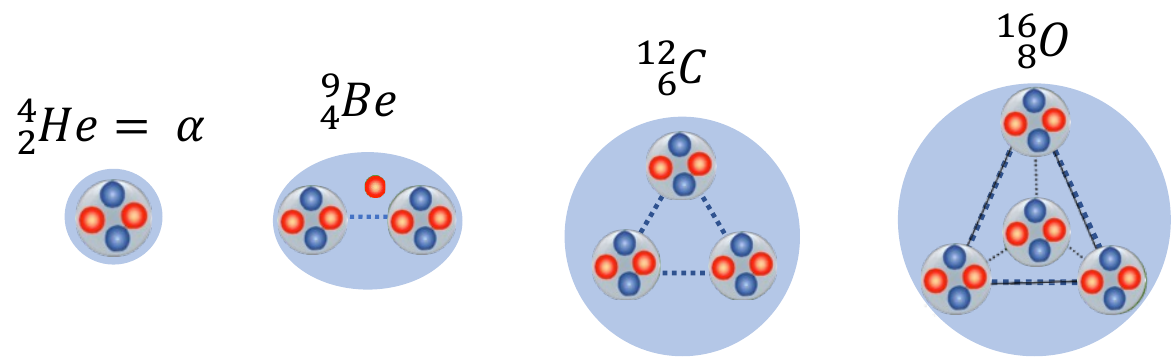}
\vskip -0.36cm
 \caption{
 Schematic illustrations of $\alpha$ clustering in atomic nuclei, panel (a) $^{4}$He=$\alpha$ particle, (b) $^{8}$Be, (c) $^{12}$C and (d) $^{16}$O. The blue areas represent the space existence of clusters in the nucleus.
 \label{fig:0}
 }
 }
 \end{figure}

Alpha clustering, commonly observed in excited states of light nuclei, extends to ground states, particularly in nuclei situated distant from the stability line~\cite{VONOERTZEN200643}. These nuclei are often depicted as quasi-molecular entities composed of clusters~\cite{R1, Ikeda:1968, Avakov:1972mlq, Wefelmeier:1937, Wheeler:1937zza, Morinaga:1956zza, He:2014iqa, Broniowski:2013dia, Zhang:2017xda, Li:2020vrg} as illustrated in Fig. \ref{fig:0}. Notably, extensive research over the past five decades has focused on elucidating 3-alpha and 4-alpha clustered structures in $^{12}$C and $^{16}$O, respectively \cite{Cook:1957zz, Fujiwara:1980, Freer:2014qoa, Schuck:2018ftf}. Antisymmetrized Molecular Dynamics (AMD) calculations have also provided insight, indicating that states above the decay threshold (such as the Hoyle band) exhibit a pronounced cluster structure. However, it's becoming evident that even within the ground state, the influence of cluster structures may not be negligible \cite{Kanada-Enyo:2006rjf, Kravvaris:2017nyj}. Nuclear lattice simulations have similarly addressed this phenomenon, further highlighting the potential significance of cluster structures in the ground state \cite{Epelbaum:2012qn}.

{\color{black}
 The concept of $\alpha$-clustering has also been explored within the context of relativistic nucleus-nucleus collisions. In these high-energy collisions, the structural information of the colliding nuclei is reflected in the properties of the resulting quark-gluon plasma (QGP)~\cite{Broniowski:2013dia, Zhang:2017xda, Rybczynski:2017nrx, Guo:2019sek, Lim:2018huo, Rybczynski:2019adt, Wang:2021ghq, Li:2020vrg, Behera:2021zhi, Ding:2023ibq, Summerfield:2021oex, Huang:2023viw, Schenke:2020mbo, Nijs:2021clz, Liu:2023gun, Giacalone:2024luz, Zhang:2024vkh, Prasad:2024ahm}. Initially, the idea was to collide a light nucleus with a heavy nucleus at high energies to understand better the $\alpha$-cluster configurations within the light nucleus~\cite{Broniowski:2013dia}. Significant efforts have been devoted to studying the effects of $\alpha$-clustering on flow harmonics and other observables in $^{16}$O+$^{16}$O collisions using various initial geometry models, hydrodynamic models, and transport models~\cite{Lim:2018huo, Rybczynski:2019adt, Wang:2021ghq, Li:2020vrg, Behera:2021zhi, Ding:2023ibq, Summerfield:2021oex, Huang:2023viw, Schenke:2020mbo, Nijs:2021clz, Epelbaum:2013paa, Rybczynski:2017nrx, Pieper:2002ne, Lee:2008fa, He:2014iqa}. In this work, we build on these previous efforts by extending the study to eA collisions at the Electron-Ion Collider (EIC).
 }

In the 2030s, the EIC will pioneer nuclear deep inelastic scattering in collider kinematics, unlocking a wealth of new opportunities for nuclear physics research~\cite{AbdulKhalek:2021gbh}. In the context of $e$+$A$ collisions at the EIC, the proposed reaction mechanism divides nuclear reactions into three distinct stages~\cite{Chang:2022hkt}: (i) hard scattering processes~\cite{Piller:1995kh}, (ii) IntraNuclear Cascade (INC) processes~\cite{Cugnon:1982qw}, and (iii) the breakup of excited nuclear remnants~\cite{Weisskopf:1937zz}. These stages are assumed to occur sequentially on different time scales in the nuclear rest frame. Hard scattering processes happen almost instantaneously with the initial collisions, while INC processes and the subsequent breakup of excited nuclei occur on the order of \(10^{-22}\) and \(10^{-16}\) seconds, respectively~\cite{Serber:1947zz, Monira:2023}. Furthermore, the time scales of these processes are expected to depend on various properties of the target nuclei and the specific collision kinematics. Consequently, the ability to isolate kinematic regions dominated by each process will enable us to refine nuclear excitation models and gain deeper insights into the internal structure of the target nucleus~\cite{Mathews:1982zz}.

In this study, we are raising the question, \textit{can the future EIC provide a deep understanding of the $\alpha$-clustering in the nuclei ground state and help constrain different theoretical models}? To provide an answer to such a question, we employed the BeAGLE model \cite{Chang:2022hkt}, a leading $e$+$A$ event generator, to study the $\alpha$-clustering in light ions at the EIC. Using the BeAGLE model framework, we explored the sensitivity of the mean energy ($\langle E \rangle$) at forward rapidity to the $\alpha$-clustering in light ions. 
In $e$+$A$ collisions, we anticipate the particles at forward rapidity to be created in various processes, such as hard collisions, nuclear evaporation, nuclear fission, and intra-nuclear cascade. In addition, we expect the energy of those particles to depend on the impact parameter and, therefore, on the nuclear configuration \cite{Chang:2022hkt}.

Here, an important objective is to develop a more strict constraint for measuring $\alpha$-clustering at the EIC using the $e$+$^{9}$Be, $e$+$^{12}$C, and $e$+$^{16}$O collisions. The paper is organized as follows. Section~\ref{sec:2} summarizes the theoretical model used to investigate the $\alpha$-clustering effect on the mean energy of the neutrons at forward rapidity. The results from the model studies are presented in Sec.~\ref{sec:3} followed by a summary in Sec.~\ref{sec:4}.
\section{Methodology}\label{sec:2}
\begin{figure}[!h]
 \centering{
 \includegraphics[width=0.99\linewidth,angle=0]{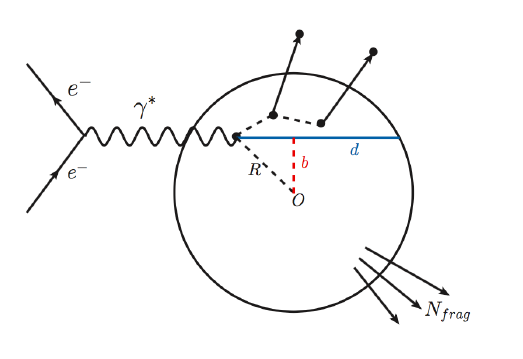}
\vskip -0.36cm
 \caption{
 Diagram illustrating geometric features of the leptonuclear scattering process in the BeAGLE model. The variable $b$ is referred to as the impact parameter, with small values indicating “central collisions” and large values indicating “peripheral collisions.” 
 \label{fig:01}
 }
 }
 \end{figure}

The study employed version 1.03 of the Monte Carlo code "Benchmark eA Generator for LEptoproduction" (BeAGLE)~\cite{Chang:2022hkt}. BeAGLE, a versatile FORTRAN program tailored for simulating electron-nucleus (eA) interactions {\color{black}as described in Fig~.\ref{fig:01}}, served as the cornerstone of our analysis. Notably, we leveraged the GLISSANDO~\cite{BOZEK2019106850, Rybczynski:2017nrx} module within BeAGLE, which we updated in the current version to explore the nuclear structure effect on the $\langle E \rangle$ measured at forward rapidity.

\subsection{The BeAGLE model}

BeAGLE operates as a hybrid model, synergizing various established codes DPMJet~\cite{Roesler:2000he}, PYTHIA6~\cite{Sjostrand:2006za}, PyQM~\cite{Dupre:2011afa}, FLUKA~\cite{Bohlen:2014buj, Ferrari:2005zk}, and LHAPDF5~\cite{Whalley:2005nh} to comprehensively describe high-energy leptonuclear scattering phenomena.
\begin{itemize}
\item{DPMJet model defines hadron creation and interactions with the nucleus via an intra-nuclear cascade.}
\item{PYTHIA-6 model, which gives the partonic interactions and subsequent fragmentation process.}
\item{PyQM model provides the geometric density distribution of nucleons in a nucleus. In addition, the model executes the Salgado-Wiedemann quenching weights to represent the partonic energy loss~\cite{SW:2003}.}
\item{FLUKA model describes the decay of the excited nuclear remnant (i.e., nucleon and light ion evaporation, nuclear fission, Fermi breakup of the decay fragments, and photon emission de-excitation). }
\item{LHAPDF5 model and FLUKA model define the high-energy lepto-nuclear scattering.}
\end{itemize}
In addition, BeAGLE provides steering, multi-nucleon scattering (shadowing), and an improved description of the Fermi momentum distributions of nucleons in the nuclei.
%

BeAGLE offers many options to manipulate simulation phenomena. This includes the ability to describe nuclear shadowing via various approaches, account for hadron formation time in the DPMJet intranuclear cascade, and tailor the Fermi motion of nucleons within the nucleus through different mechanisms.  PyQM functionalities extend to specifying the transport coefficient $\hat{q}$ to modulate the interaction degree between energetic partons and the nuclear environment and fine-tune details of the partonic energy loss process. The main program, DPMJet, interfaces with PYTHIA6 to handle elementary interactions and fragmentation. PyQM, in turn, manages this process directly post-elementary interactions in PYTHIA6, while DPMJet is responsible for nuclear geometry and nuclear evaporation post-fragmentation facilitated by FLUKA.

\begin{figure}[!h]
\centering{
\includegraphics[width=0.99\linewidth,angle=0]{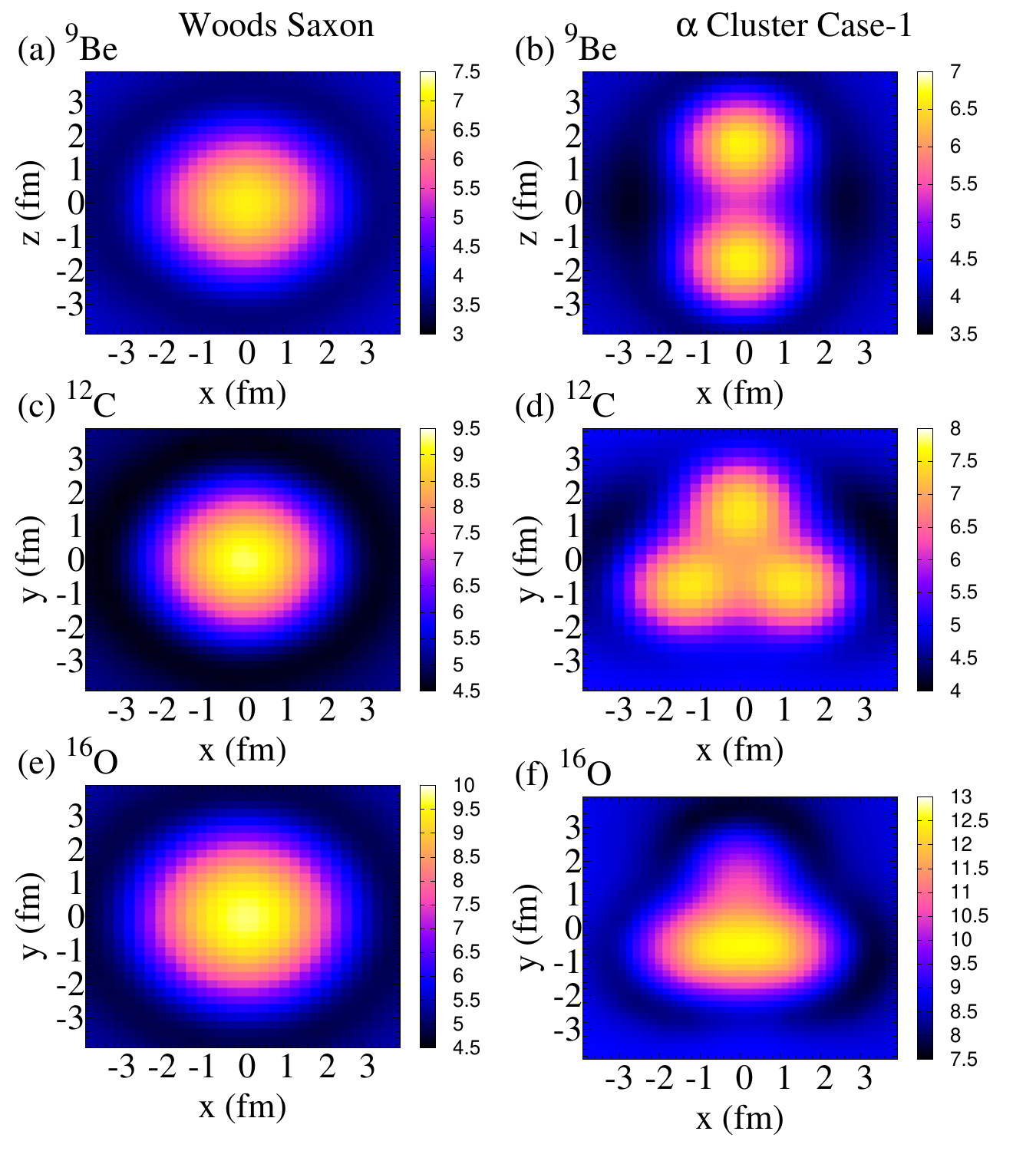}
\caption{
Comparison of the nucleons density distribution at different positions generated using
the Woods Saxon Distribution (a),(c),(e) and the $\alpha$-clustering model {\color{black}with fixed orientation along the z-axis (b),(d),(f) for $^{8}$Be, $^{12}$C, and $^{16}$O.}
\label{fig:1}
}
}
\end{figure}

The BeAGLE model, as it currently stands, omits clustering and cluster configurations. Therefore, we adopt the $\alpha$-clustering provided by the GLISSANDO model prescription~\cite{BOZEK2019106850,Rybczynski:2017nrx} for $^{9}$Be (2$\alpha$+neutron), $^{12}$C (3$\alpha$), and $^{16}$O (4$\alpha$) (see Fig.~\ref{fig:1}). In the GLISSANDO model, as illustrated in Fig.~\ref{fig:1}, the centers of the clusters are separated by a distance $d$. The nucleons' position in each cluster is randomly chosen according to a Gaussian distribution;
\begin{eqnarray}
f_{i}(\vec{r}) = A~e^{-\frac{3}{2}  \left( \frac{\vec{r}-\vec{c_{i}}}{\sigma_{c}}   \right)^{2}  }, 
\end{eqnarray}
where $\vec{r}$ is the x, y and z coordinate of a nucleon, $\vec{c_i}$ is the $i^{th}$ cluster center, and $\sigma_{c}$ is the RMS radius of the cluster. Then, the nucleons' x, y, and z positions are generated successively, swapping between clusters. 
In the current study, the size of the $\alpha$-cluster is 1.1~fm for all systems, and the distance between the centers of clusters is chosen to be 3.6, 2.8 a.d 3.2~fm for  $^{9}$Be, $^{12}$C, and $^{16}$O respectively. Note that for $^{9}$Be, we have an extra neutron in addition to the two $\alpha$ clusters. The extra neutron is included via the distribution;
\begin{eqnarray}
f_n(\vec{r})=B~r^2~e^{-\frac{3}{2}  \left( \frac{r}{\sigma_{n}}   \right)^{2} }.
\end{eqnarray}
In addition, the centers of each nucleon pair are forbidden from being closer than the expulsion distance of 0.9~fm~\cite{Broniowski:2010jd}.
Finally, the obtained distributions are shifted such that their center of mass is positioned at the origin of the coordinate frame. Consequently, we get the Monte Carlo distributions with the cluster correlations included.

{\color{black}
In this work, we are comparing the results evaluated via the $\alpha$-clustering  scheme described in this work with the results assessed via Woods Saxon Distribution given by the standard GLISSANDO model~\cite{BOZEK2019106850, Rybczynski:2017nrx}. In the GLISSANDO model, the Woods-Saxon distribution is given as follows:
\begin{eqnarray}
\rho(r) = c \frac{4 \pi r^{2}}{1 + exp((r-R)/a)},
\end{eqnarray}
where the constant c is such a normalization, and the parameters R and a can be found in Ref.~\cite{DeVries:1987atn}. 

}

In the present work, nucleons in a nucleus are distributed according to four cases;
\begin{itemize}
\item{Case-1, according to the Woods–Saxon density distribution.}
\item{Case-2, according to our description for clustered nuclei with a dipole, trigonal planar, and tetrahedral shapes for $^{9}$Be, $^{12}$C and $^{16}$O respectively. In Case-2, the clustered nuclei have a fixed orientation along the z-axis.}
\item{Case-3, same as case-2 but with random orientation.}
\item{ {\color{black}Case-4, According to our description for clustered nuclei but with all three or four $\alpha$ are lined up in a straight line~\cite{Huang:2017ysr} for $^{12}$C and $^{16}$O respectively. In this case, the $^{12}$C and $^{16}$O nuclei are allowed to rotate randomly (i.e., random orientation).} }
\end{itemize}
{\color{black}
It's important to point out that we do not consider mixed cases between clustering and no clustering effects.
}

\subsection{Observable}

{\color{black}
The BeAGLE model identifies three reaction stages in the context of $e$+$A$ collisions: hard scattering processes, INC processes~\cite{Cugnon:1982qw}, and the breakup of excited nuclei~\cite{Serber:1947zz, Monira:2023}. The hard scattering processes occur too rapidly to be significantly influenced by the details of the nuclear structure, and the nuclear breakup stage primarily reflects the properties of the excited remnant, with no memory of the prior stages~\cite{Ferrari:2005zk}. Consequently, within the BeAGLE framework, neither stage is suitable for studying the nuclear structure. In contrast, the INC processes, triggered by particles produced at the primary interaction following a straight-line trajectory and undergoing successive collisions (see Fig.~\ref{fig:01}), provide valuable information about the initial nuclear structure of the target nuclei. Consequently, we test our ability to separate the kinematic regions where each of these processes dominates by studying the $dN/d\eta$ in $e$+$A$ collisions.

\begin{figure}[!h]
\centering{
\includegraphics[width=0.99\linewidth,angle=0]{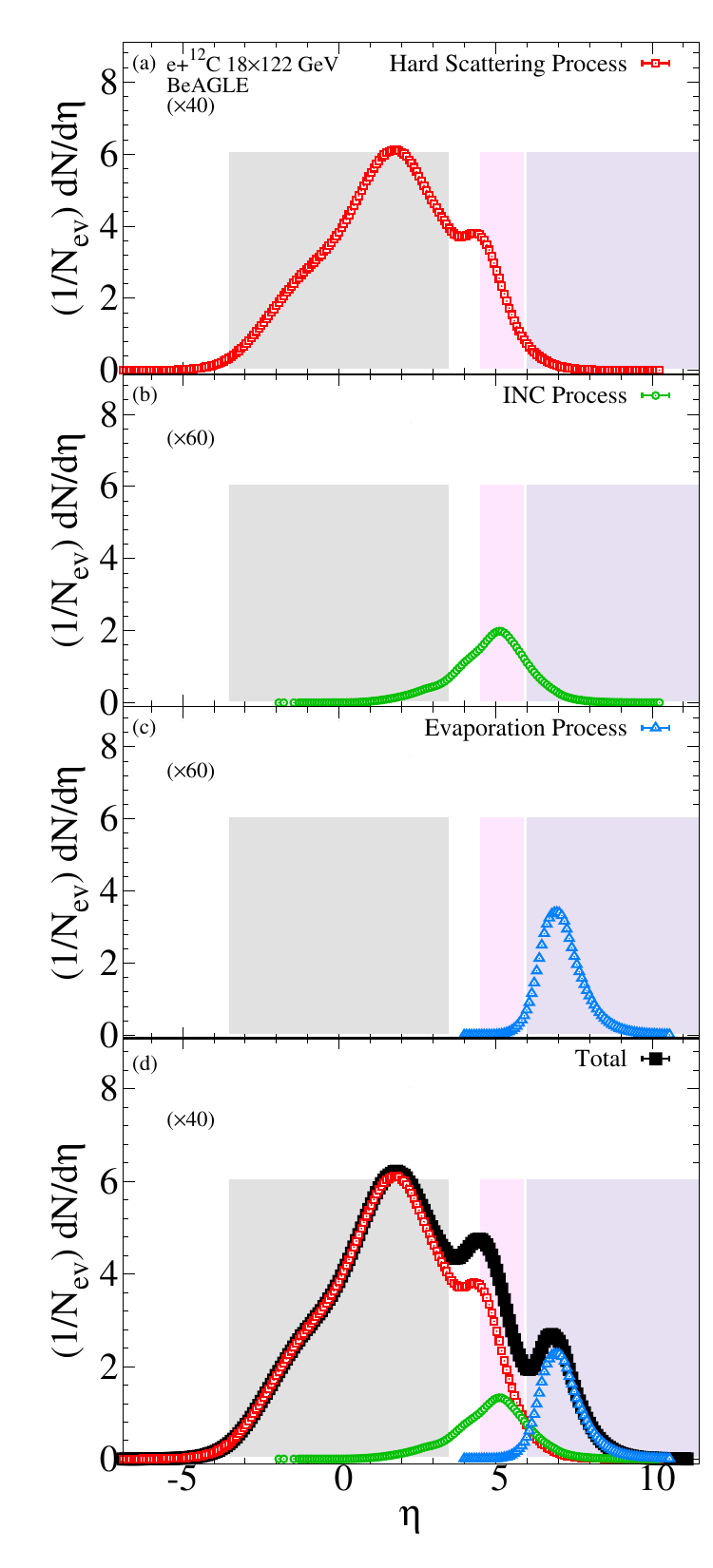}
\caption{
Comparison of {\color{black}the $\eta$ distributions of the normalized charge hadrons multiplicity}  ($dN/d\eta$) produced in hard scattering process panel (a), INC process panel (b), evaporation process panel (c), and the total sum of the three stages in panel (d) for e+$^{12}C$ at 18$\times$122 GeV. The bands represent the $\eta$ regions given by the $ePIC$~\cite{ePIC} detector acceptance.
\label{fig:1x}
}
}
\end{figure}

{\color{black} Figure~\ref{fig:1x} shows the {\color{black}the $\eta$ distributions of the normalized charge hadrons multiplicity} ($dN/d\eta$ ) created in the hard scattering process, the INC process, and the evaporation process for $e$+$^{12}$C at 18$\times$122 GeV. Our results indicate that these three stages are dominated at mean values of $1.8$, $4.9$, and  $6.9$  for the mid-, forward-, and far-forward rapidity, respectively.
In this work, the $\eta$ selection is guided by the expected acceptance of the $ePIC$ detector~\cite{ePIC}. The three bands highlighted in Fig.~\ref{fig:1x} correspond to the acceptance ranges of the three primary subsystems anticipated for the $ePIC$ detector. These subsystems are expected to cover the rapidity ranges $|\eta| < 3.5$ for mid-rapidity, $4.5 < \eta < 5.9$ for forward rapidity, and $\eta > 6.0$ for far-forward rapidity.
}  Consequently, we chose to study the mean energy of particles ($\langle E \rangle$) at forward rapidity $4.5 < \eta < 5.9$ as our main observable. 
The $\langle E \rangle$ is given as:
\begin{eqnarray}
     \langle E \rangle = \frac{\sum_{i} w_{i} E_{i}}{\sum_{i} w_{i}},
\end{eqnarray}
where, the sum runs over the particle in the $\eta$ acceptance between $4.5$--$5.9$,  $E_{i}$ and $w_{i}$ are the i$^{th}$ particle energy and wight. For a perfect detector $w_{i} = 1.0$. {\color{black}
In future work, we will consider the expected detector efficiency of the ePIC detector~\cite{ePIC}.
}

This work presents the results as functions of collision centrality, defined via cuts over the impact parameter distribution (see Fig.~\ref{fig:01}). 

\subsection{Results and discussion}\label{sec:3}
Before delving into nuclear structure studies at the forthcoming EIC, we meticulously prepared the initial coordinate space for the light nuclei targets, as depicted in Fig.~\ref{fig:1}. We then utilized the BeAGLE model framework to generate simulated data under standardized conditions for each collision system. Our approach began with a focused analysis of scenarios where the nuclei are oriented in fixed positions. We chose distinct spatial configurations for different nuclei: ${^9}$Be with two clusters aligned along the collision axis (z-axis), ${^{12}}$C with three clusters situated within the plane perpendicular to the collision axis (x-y plane), and ${^{16}}$O with three clusters within the x-y plane and an additional cluster along the z-axis. This preliminary investigation aims to ensure the correct implementation of the initial nuclear configuration into the BeAGLE model.

\begin{figure*}[!h]
\centering{
\includegraphics[width=0.70\linewidth,angle=0]{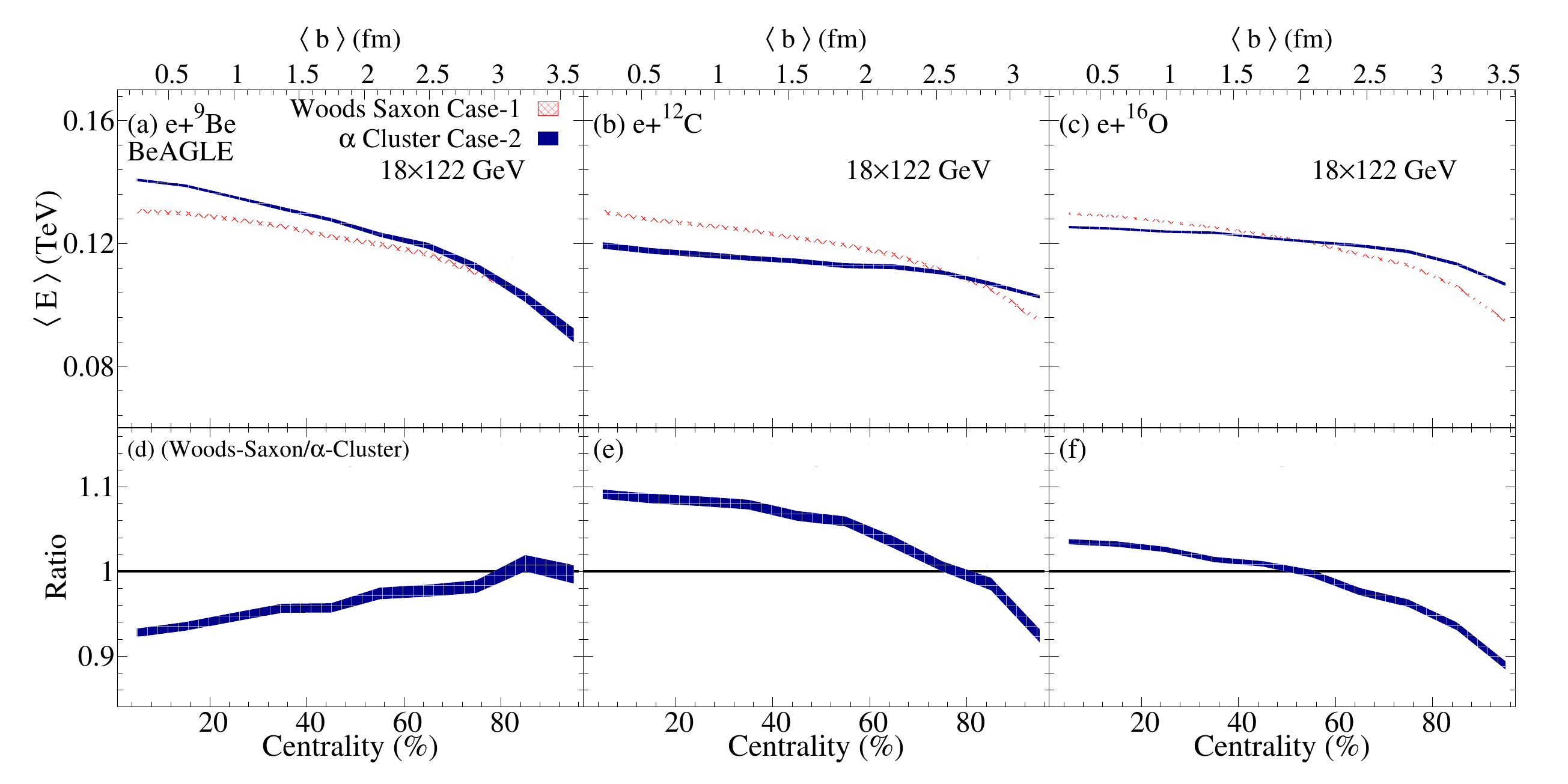}
\vskip -0.30cm
\caption{
The centrality and impact parameter dependence of the mean energy of the particles in $4.6 < \eta < 5.9$ for $e$+$^{9}$Be (a), $e$+$^{12}$C (b) and $e$+$^{16}$O (c) from the BeAGLE model. Panels (d), (e), and (f) show the ratios between the Woods–Saxon density distribution Case-1 and $\alpha$ clustering with fixed orientation nuclei Case-2.
~\label{fig:3}
}
}
\end{figure*}
\begin{figure*}[!h]
\centering{
\includegraphics[width=0.70\linewidth,angle=0]{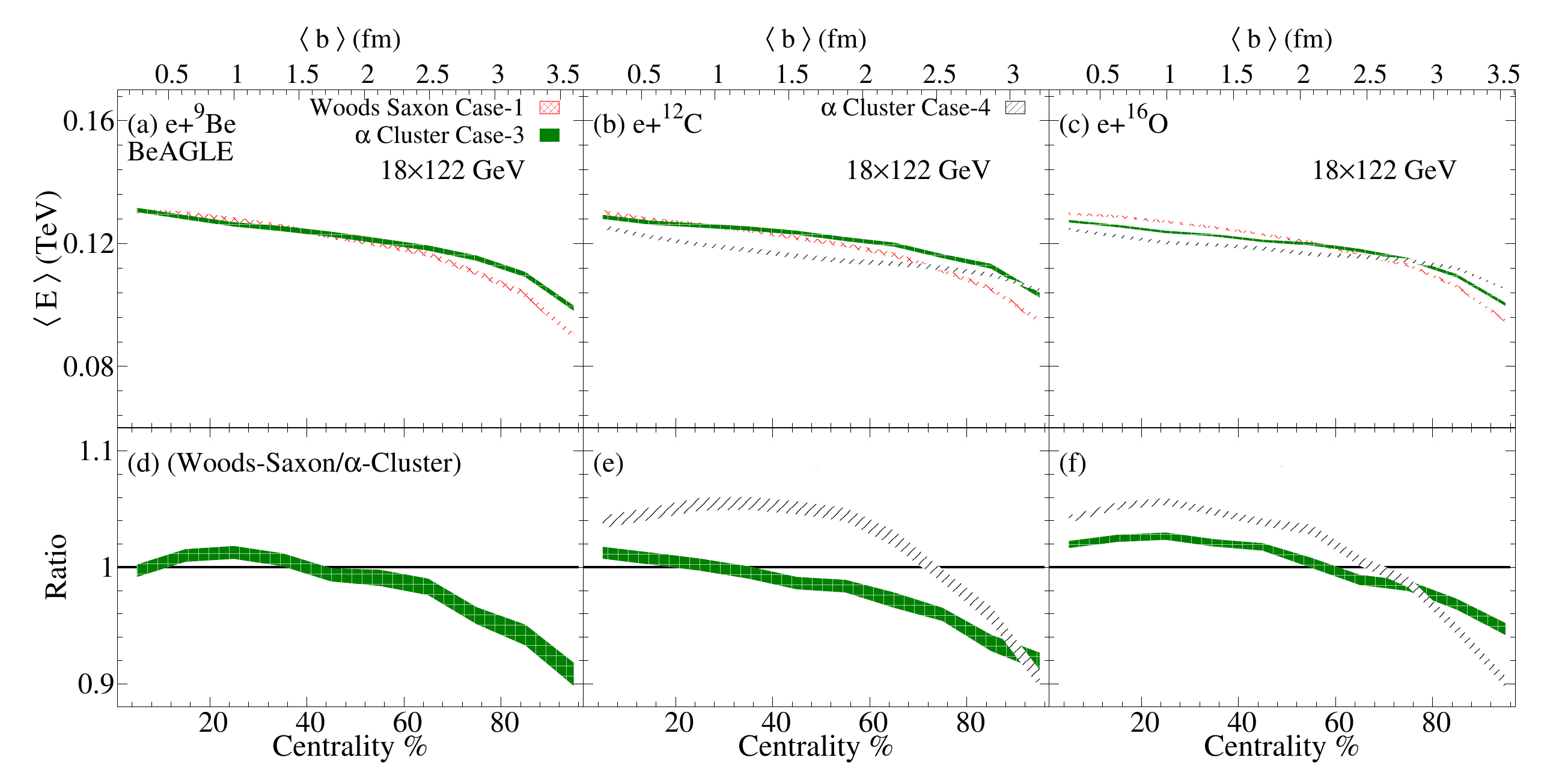}
\vskip -0.30cm
\caption{
Same as in Fig.~\ref{fig:3} but for clustering with random orientation nuclei Case-3 and Case-4.
~\label{fig:4}
}
}
\end{figure*}
\begin{figure*}[!h]
\centering{
\includegraphics[width=0.70\linewidth,angle=0]{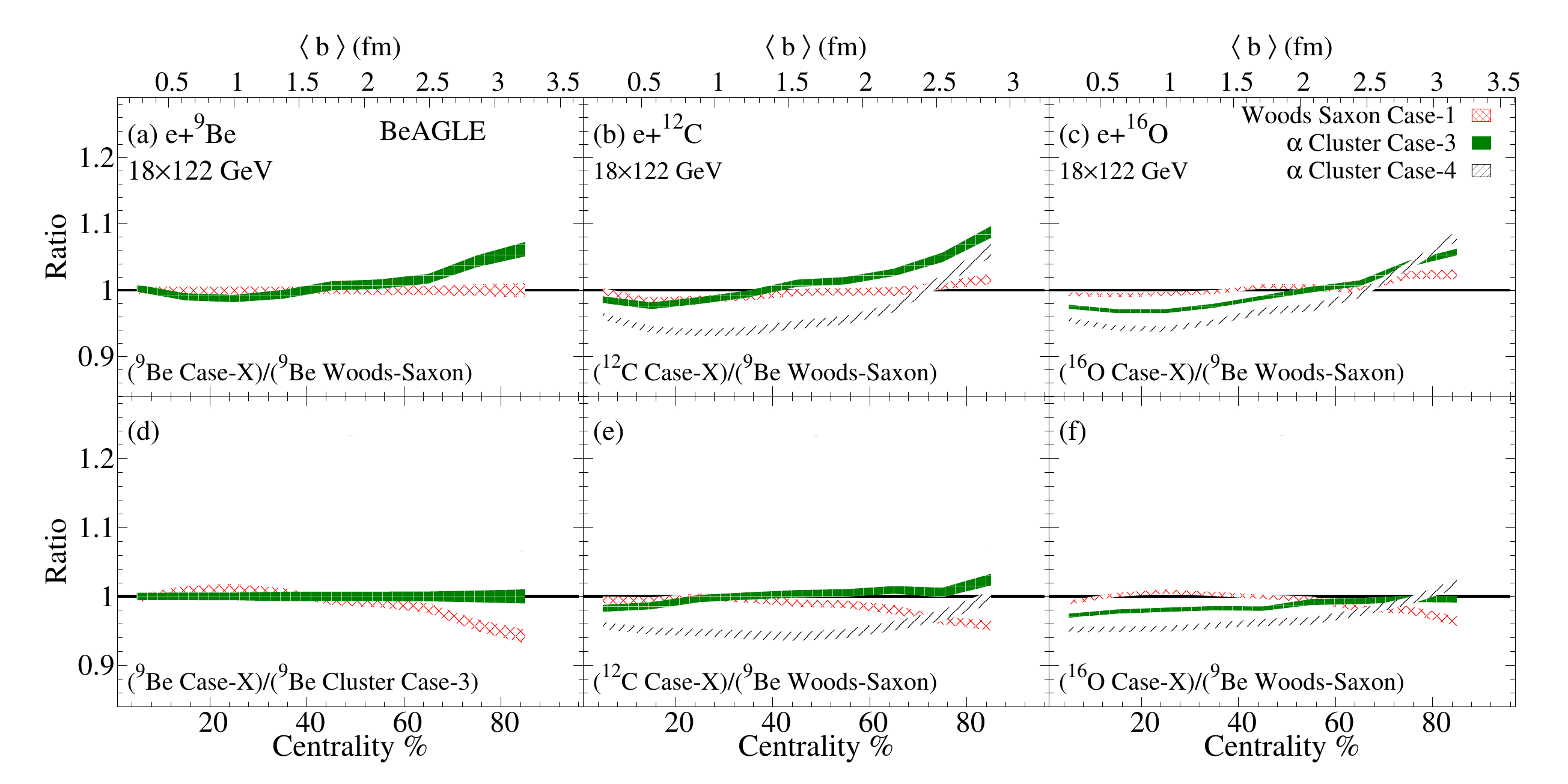} 
\vskip -0.30cm
\caption{
The centrality and impact parameter dependence of the $\langle E \rangle$ ratios to the $^{9}$Be Case-1 panels (a), (b) and (c), and to the $^{9}$Be Case-3 panels (e), (f) and (g).
~\label{fig:5}
}
\vskip -0.30cm
}
\end{figure*}

Figure~\ref{fig:3} panels (a), (b), and (c) provide a comparative analysis of the average energy ($\langle E \rangle$) of particles within the pseudorapidity range $4.6 < \eta < 5.9$ for $e$+$^{9}$Be, $e$+$^{12}$C, and $e$+$^{16}$O collisions, respectively, utilizing data generated by the BeAGLE model under fixed orientation nuclei conditions. Our findings across all collision systems consistently revealed a decrease in $\langle E \rangle$ as collision centrality transitioned towards more peripheral configurations.
Further insights into the impact of different nuclear configurations on particle energies were gleaned by comparing $\langle E \rangle$ ratios between scenarios with Woods–Saxon density distribution (Case-1) and those with $\alpha$ clustering under fixed orientation nuclei conditions (Case-2), as depicted in Fig.~\ref{fig:3} panels (d), (e), and (f). In the case of $e$+$^{9}$Be collisions (panel (d)), where two alpha clusters align along the collision axis, we observed a higher $\langle E \rangle$ for the $\alpha$ cluster Case-2 compared to the Woods–Saxon Case-1 in central collisions. Conversely, for $e$+$^{12}$C collisions (panel (e)), featuring three alpha clusters within the x-y plane, there was a reduction in $\langle E \rangle$ for the $\alpha$ cluster Case-2 relative to the Woods–Saxon Case-1 in central to mid-central collisions. Notably, in $e$+$^{16}$O collisions (panel (f)), characterized by three alpha clusters in the x-y plane and one cluster along the z-axis, we observed a reduction in $\langle E \rangle$ for the clustered Case-2 in central collisions and an increase in peripheral collisions.
Our analyses underscore the sensitivity of $\langle E \rangle$ at forward rapidity to the underlying $\alpha$ clustering structure, as evidenced by the presented ratios. 

In practice, the orientation of target nuclei cannot be fixed, necessitating a reevaluation of our analysis with randomized nuclei orientations. Similar to the methodology employed for Fig.~\ref{fig:3}, we present a comparative study in Fig.~\ref{fig:4} panels (a), (b), and (c), illustrating the average energy ($\langle E \rangle$) of particles within the pseudorapidity range $4.6 < \eta < 5.9$ for $e$+$^{9}$Be, $e$+$^{12}$C, and $e$+$^{16}$O collisions, respectively, utilizing data from the BeAGLE model. Across all scenarios, our computations consistently reveal a reduction in $\langle E \rangle$ as collision centrality progresses towards more peripheral selections.
The corresponding $\langle E \rangle$ ratios relative to the Woods–Saxon distribution (Case-1) are depicted in panels (d), (e), and (f) of Fig.~\ref{fig:4}. Notably, our findings demonstrate a consistent trend among the various collision systems. Specifically, we observe an increase in $\langle E \rangle$ for the clustered case in peripheral collisions, with the magnitude of this increase depending on the system size and the configuration of clusters (e.g., chain configuration). These observations underscore the sensitivity of $\langle E \rangle$ at forward rapidity to the $\alpha$ clustering structure and configurations, emphasizing the importance of considering nuclear structure effects in our analyses.

Furthermore, we expanded our analysis by constructing a series of ratios between (i) each system and the $e$+$^{9}$Be Case-1 and (ii) each system and the $e$+$^{9}$Be Case-3. These ratios offer insights into the influence of nuclear structure on energy distributions. Ideally, these ratios should approximate unity if no discernible nuclear structure is present, implying that all systems can be adequately described by a simple Woods–Saxon distribution. Conversely, deviations from unity in these ratios would suggest the presence of nuclear structure effects.

Figure~\ref{fig:5} illustrates the dependence of the $\langle E \rangle$ ratios on the $^{9}$Be Case-1 (panels (a), (b), and (c)) and the $^{9}$Be Case-3 (panels (d), (e), and (f)). Our analysis revealed that for Case-1, where nuclear structure effects are absent, all ratios closely approximate unity. However, when the ratios incorporate systems with included nuclear structure, deviations from unity become apparent.  This observation reinforces the importance of accounting for nuclear structure effects when interpreting energy distributions in $e$+$A$ collisions.


\section{Future investigations}\label{sec:4}
In this work, we investigated the influence of the nuclear structure of light nuclei on the final state particles produced at forward rapidity. Our investigation encountered two key aspects summarized as follows:
\begin{itemize}
    \item Centrality definition in e+A collisions:\\
    The $\langle E \rangle$ results are presented as functions of collision centrality, defined by cuts over the impact parameter distribution. In e+A collisions, the impact parameter is independent of collision kinematics and weakly depends on the final state particles~\cite{Chang:2022hkt, Zheng:2014cha}, as discussed in Appendix~\ref{APX-1}. These properties complicate the identification of a final state measure strongly correlated with the impact parameter.

    \item Sensitivity of $\langle E \rangle$ Correlators:\\
    The ratios presented in Figs.~\ref{fig:4} and \ref{fig:5} indicate a maximum sensitivity of approximately 10\% to the differences in nuclear structure between the presented cases and systems. Our observations raise the question of whether such differences can be experimentally measured. Answering this question requires detailed consideration of detector specifics and the machine's capabilities to handle different collision systems, which exceeds the scope of our current analysis.
\end{itemize}
We view this paper as an important step towards understanding the nuclear structure of light nuclei in the context of e+A collisions. However, finding creative ways to (i) define centrality in e+A collisions, (ii) identify new, centrally independent correlators, and (iii) assess whether a 10\% difference can be experimentally detected are crucial aspects that will be addressed in future work.

{\color{black}
Here, it's also important to point out that nuclear structure (i.e., clustering) is a low-resolution phenomenon that fundamentally informs us that there are significant spatial correlations of nucleons in the nuclear ground state. Consequently, such correlations can be probed via observables sensitive to multi-nucleon correlations in the target nucleus~\cite{Mantysaari:2023qsq}.  In the future, one may consider replacing the $\langle E \rangle$ with more differential correlators that should be invented and investigated to pinpoint nucleon-correlation effects in eA collisions.
}

\section{Conclusion}\label{sec:4}
Using the modified BeAGLE model framework, we explored the potential of studying the effect of nuclear structure (i.e., $\alpha$ clustering) on the produced final state particles at forward rapidity. Our $\langle E\rangle$ results of light nuclei indicated centrality characteristic patterns that depend on the clusters' configurations and the system size. In addition, we constructed a set of ratios between different systems that deviate from unity when one or more of the systems contain nuclear structures. 
Our findings underscore the sensitivity of both $\langle E \rangle$ and system size ratios of particles at forward rapidity to the presence and configurations of $\alpha$ clustering within the nucleus. This sensitivity underscores the intricate interplay between nuclear structure and particle production dynamics in e+A collisions, emphasizing the need to consider these effects in comprehensive analyses.
Consequently, our results lend credence to the proposal put forth by antisymmetrized molecular dynamics and nuclear lattice simulations regarding the potential presence of cluster structures in the ground state of light nuclei. The validation of these hypotheses at the forthcoming Electron-Ion Collider (EIC) holds promise for advancing our understanding of nuclear structure and its implications for particle production mechanisms.

\section*{Acknowledgments}
This work was supported in part by funding from the Division of Nuclear Physics of the U.S. Department of Energy under Grant No. DE-FG02-96ER40982 (NM) and DE-SC0012704 (ZT). ZT acknowledges the support of the Laboratory Directed Research and Development (LDRD) 22-027 and LDRD-23-050 project. AR acknowledges the support of the American University in Cairo under the agreement Number SSE-PHYS-A.H-FY24-RG-2023-Dec-17-14-59-41.



\appendix
\section{Centrality definition in e+A collisions}
\label{APX-1}
In prior investigation~~\cite{Chang:2022hkt, Zheng:2014cha}, it has been proposed that the energy deposited in the Zero-Degree Calorimeter (ZDC) with angular acceptance $\theta$ $<$ $5.5$ rad ($\eta$ $>$ 6.0) can be used to define the centrality in e+A collisions. Figure~\ref{fig:Ax1} presents the correlations between energy deposited in the ZDC and the impact parameter in panel (a). Such correlations indicated that the ZDC deposited energy increases weakly as the b decreases. Figure~\ref{fig:Ax1} panel (b) shows the b distributions in central (0\%–1\%) and peripheral (60\%–100\%) collisions. The distributions are normalized by the total number of events. A weak difference is observed between central and peripheral collisions. Such observation suggests that using the ZDC energy as an experimental handle on the collision geometry is complicated.

\begin{figure}[!h]
\centering{
\includegraphics[width=0.99\linewidth,angle=0]{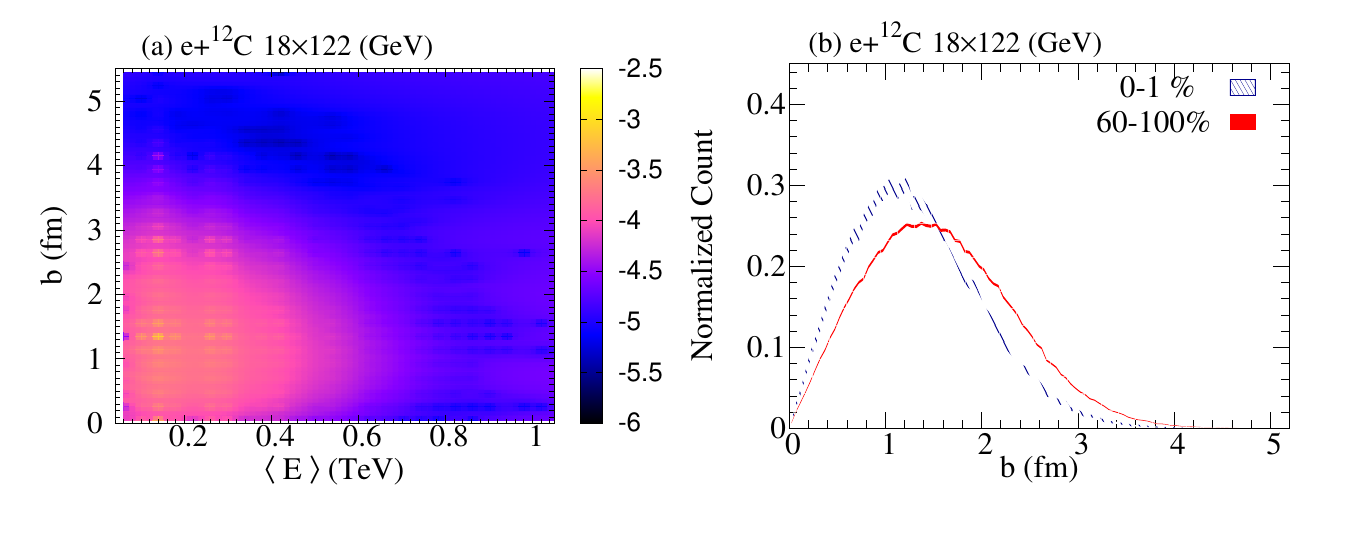}
\vskip -0.30cm
\caption{
The correlations between energy deposited in the Zero-Degree Calorimeter (ZDC) and the impact parameter (b) is shown in panel (a).  The comparison of b distributions in central and peripheral collisions as determined by Zero-Degree Calorimeter energy cuts are presented in panel (b).}
~\label{fig:Ax1}
}
\end{figure}


\bibliography{ref}

\end{document}